\documentclass[%
 reprint,
superscriptaddress,
 aps,
 prl,
]{revtex4-1}

\usepackage{graphicx}
\usepackage{dcolumn}
\usepackage{bm}
\usepackage{natbib}
\usepackage{amsmath,bm}
\usepackage{amssymb}
\usepackage{hyperref}
\usepackage{siunitx}
\usepackage[a4paper,margin=2cm]{geometry}
\usepackage[caption=false]{subfig}
\usepackage{booktabs}
\usepackage[export]{adjustbox}
\usepackage{multirow}
\usepackage{physics}
\usepackage{accents}

\usepackage{multibib}
\usepackage[dvipsnames]{xcolor}
\newcommand{\zm}[1]{\textcolor{black}{#1}}

\begin{document}

\title{Mutual friction and diffusion of two-dimensional quantum vortices}

\author{Zain Mehdi}
 \email{zain.mehdi@anu.edu.au}
 \affiliation{Department of Quantum Science and Technology and Department of Fundamental and Theoretical Physics, Research School of Physics, Australian National University, Canberra 2600, Australia}%
\author{Joseph J. Hope}
\affiliation{Department of Quantum Science and Technology and Department of Fundamental and Theoretical Physics, Research School of Physics, Australian National University, Canberra 2600, Australia}%
\author{Stuart S. Szigeti}
\affiliation{Department of Quantum Science and Technology and Department of Fundamental and Theoretical Physics, Research School of Physics, Australian National University, Canberra 2600, Australia}%
\author{Ashton S. Bradley}
\affiliation{The Dodd-Walls Centre for Photonic and Quantum Technologies, Department of Physics, University of Otago, Dunedin, Aotearoa New Zealand
}

\date{\today}

\begin{abstract}
We present a microscopic \zm{open quantum systems} theory of thermally-damped vortex motion in oblate atomic superfluids that \zm{includes} previously neglected \zm{energy-damping} interactions between superfluid and thermal atoms. \zm{This mechanism couples strongly to vortex core motion and causes dissipation of vortex energy} due to mutual friction, as well as Brownian motion of vortices due to thermal fluctuations. We \zm{derive} an analytic expression for the dimensionless mutual friction coefficient that gives excellent quantitative agreement with experimentally measured values, without any fitted parameters. \zm{Our work closes an existing two orders of magnitude gap between dissipation theory and experiments, previously bridged by fitted parameters, and provides a microscopic origin for the mutual friction and diffusion of quantized vortices in two-dimensional atomic superfluids.} 
\end{abstract}


\maketitle

Quantum vortex dynamics are central to many superfluid phenomena, including the Kibble-Zurek mechanism \cite{Zurek2005}, BKT transition \cite{Kosterlitz_1973,Berezinsky1972}, persistent current decay \cite{Mathey2014,rooney_persistentcurrent_2013}, and type-II superconductivity~\cite{Abrikosov1957}. 
\zm{Planar atomic superfluids offer an ideal platform for studying non-equilibrium superfluid behaviour, with dynamics significantly simplified by forcing vortex lines to align with the tightly confined axis and move as points in the plane~\cite{fetter_vortices_1966} in perfect analogy to two-dimensional (2D) electrodynamics~\cite{Ambegaokar1980}. Experimentally, these systems have demonstrated powerful capabilities for studying incompressible `point-vortex' turbulence~\cite{gauthier_giant_2019,johnstone_evolution_2019,stockdale_universal_2020,reeves_turbulent_2022}, arguably the simplest manifestation of 2D turbulent flow~\cite{onsager_statistical_1949}.} \par 

\zm{Dissipation of vortex energy is an important aspect of the non-equilibrium dynamics of 2D quantum vortices, and is critical to spectral energy transport~\cite{billam_spectral_2015}, turbulent cascades~\cite{reeves_inverse_2013}, the emergence of negative-temperature vortex clusters~\cite{billam_onsager-kraichnan_2014,kim_role_2016,gauthier_giant_2019}, and turbulent relaxation of vortex matter~\cite{kwon_relaxation_2014,stockdale_universal_2020,reeves_turbulent_2022}. In superfluid helium, vortex dissipation is understood as arising due to `mutual friction' between moving vortices and the normal component of the fluid~\cite{hall_rotation_1956,iordanskii_mutual_1966,barenghi_friction_1983}. This approach reduces to phenomenology for weakly-interacting systems which cannot be described within a two-fluid approximation, leaving the microscopic origin of thermal dissipation of vortex energy in atomic superfluids an important open question. }\par

\zm{Although the conservative dynamics of point vortices in 2D superfluids are well-described by the Helmholtz-Kirchoff point-vortex model (PVM)~\cite{helmholtz_lxiii._1867,kirchhoff_gustav_robert_vorlesungenuber_1876}, there is currently no microscopic theory of 2D vortex dynamics that can \emph{ab initio} account for vortex dissipation observed in experiment. Existing microscopic estimates of vortex damping rates~\cite{blakie_dynamics_2008,proukakis_finite-temperature_2008} are orders of magnitude lower than observed in experiments~\cite{moon_thermal_2015,kim_role_2016,gauthier_giant_2019,stockdale_universal_2020,kwon_sound_2021,reeves_turbulent_2022}, requiring vortex damping to be treated phenomenologically as a fitted parameter~\cite{moon_thermal_2015,kim_role_2016,gauthier_giant_2019,johnstone_evolution_2019,stockdale_universal_2020,reeves_turbulent_2022}.  Furthermore, there exists no microscopic theory for recent observations of vortex diffusion driven by Brownian motion~\cite{reeves_turbulent_2022}. The lack of a complete theory of vortex damping and noise is therefore a significant barrier to establishing a strong understanding of 2D vortex dynamics and superfluid turbulence.  }

In this Letter, we present a microscopic model of vortex damping and diffusion in 2D due to thermal friction, derived from first principles reservoir theory of finite-temperature atomic Bose gases. Our approach identifies the number-conserving scattering between superfluid and thermal atoms as the dominant vortex dissipation mechanism, \zm{contrasting with previous approaches that neglect this interaction and focus instead on Bose-enhanced particle transfer between superfluid and normal fluid}. Our model includes both a dissipative mutual friction term and a stochastic term that describes the Brownian motion of vortices due to thermal fluctuations; crucially, it contains no fitted parameters and allows the mutual friction coefficient to be determined analytically from first principles. Our microscopic model's predictions are in close quantitative agreement with previous experimental measurements of the mutual friction coefficient. This establishes a microscopic justification for vortex damping phenomenology and opens the door for quantitative \emph{ab initio} modelling of two dimensional quantum turbulence (2DQT) experiments using stochastic point vortex theory. \par 

\emph{Vortex dissipation phenomenology.---}
\zm{The dynamics of dissipative 2D point-vortices in atomic superfluids is accurately described by the PVM with the addition of a phenomenological
longitudinal damping force~\cite{moon_thermal_2015,kim_role_2016,gauthier_giant_2019,stockdale_universal_2020,kwon_sound_2021,reeves_turbulent_2022}.
For a system of $N$ point vortices with positions $\mathbf{r}_i(t)=(X_i(t),Y_i(t),0)$ and unit charges $\hbar q_i/m=\pm \hbar/m$, this gives the `damped-PVM':} 
\begin{align}
\label{eq:dPVModel_Old}
\dot{\mathbf{r}}_i = \mathbf{v}_i^0 - \alpha q_i \mathbf{\hat{z}}\times\mathbf{v}_i^0  \,.
\end{align}
Here $\mathbf{v}_i^0 = \frac{\hbar}{m}\sum_{j \neq i} \frac{q_j}{r_{ij}^2} (Y_j-Y_i,X_i-X_j, 0)$ is the local superfluid velocity at the $i$-th vortex, with $r_{ij}^2 \equiv (X_i-X_j)^2 + (Y_i-Y_j)^2$, $\mathbf{\hat{z}}$ is the unit vector perpendicular to the 2D superfluid plane, and $\alpha$ is the dimensionless mutual friction coefficient. Note $\alpha=0$ gives the idealized PVM. The damped-PVM can be derived from a variational treatment of the dissipative GPE (dGPE) \cite{tornkvist_vortex_1997}, a complex Ginzburg-Landau equation where an imaginary component is added to the prefactor of the GPE. \par 

The dGPE itself can be microscopically derived from the stochastic projected GPE (SPGPE), a first-principles reservoir theory that quantitatively describes the finite-temperature dynamics of ultracold Bose gases \zm{with no fitted parameters}~\cite{blakie_dynamics_2008}. Importantly, the SPGPE provides a microscopic origin to superfluid energy dissipation due interactions between the partially-coherent modes of the quantum field (which includes the superfluid component) and the sparsely-occupied high-energy incoherent modes, which are treated as a static thermal reservoir of temperature $T$. The division between the condensed modes and reservoir is formalized by a high-energy energy cutoff that typically satisfies $\epsilon_{\rm cut}\gtrsim 2\mu$~\cite{rooney_decay_2010}, where $\mu$ is the reservoir's chemical potential~\cite{SupplementalMaterials}.

\begin{figure}
	\centering
	\includegraphics[width=1\columnwidth]{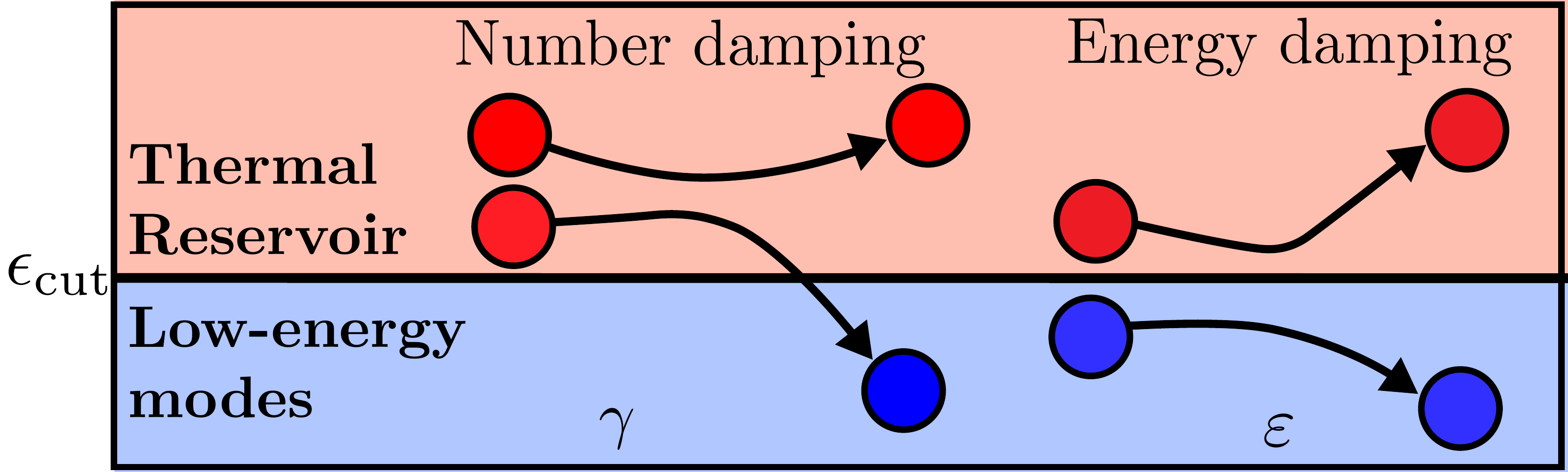}
	\caption{
	The two reservoir processes described by SPGPE theory. Number damping ($\gamma$) involves interatomic scattering that transfers atoms from the high occupation low-energy modes to the thermal reservoir and vice versa; e.g., a collision of two reservoir atoms that transfers most of the collision energy to one atom, leaving the other in a low-energy mode. Energy damping ($\varepsilon$), in contrast, describes number-conserving scattering interactions that exchange energy between the low-energy modes and reservoir without exchanging atoms.}
    \label{fig:ReservoirInteractions_Schematic}
\end{figure}

Within SPGPE theory, the dissipative prefactor in the dGPE arises due to \emph{number damping}, a reservoir process where interatomic scattering transfers atoms from the superfluid to the thermal reservoir (or vice versa) - see Fig.~\ref{fig:ReservoirInteractions_Schematic}. The mutual friction coefficient $\alpha$ in Eq.~\eqref{eq:dPVModel_Old} corresponds to the first-principles dimensionless coefficient $\gamma$~\cite{tornkvist_vortex_1997}, which is typically of order $\sim 10^{-5}$~\cite{bradley_bose-einstein_2008} - several orders of magnitude too weak to account for the mutual friction coefficients $\alpha\sim 10^{-2}$ observed in experiment~\cite{moon_thermal_2015}. Additionally, the dGPE neglects incoherent thermal fluctuations described by SPGPE theory, limiting its validity to relatively low temperatures far from the critical temperature. Despite this, phenomenological treatments of the dGPE have successfully been used to study vortex dissipation in 2DQT~\cite{bradley_energy_2012,reeves_classical_2012,reeves_identifying_2015,groszek_onsager_2016,billam_onsager-kraichnan_2014,billam_spectral_2015,reeves_turbulent_2022,kim_role_2016,stagg_generation_2015,baggaley_decay_2018}. \par 
In addition to number damping, SPGPE theory also contains a second reservoir process in which energy is exchanged due to number-conserving scattering interactions between superfluid and thermal atoms (Fig.~\ref{fig:ReservoirInteractions_Schematic}). This process is called \emph{energy damping} and, due to its computational complexity, it has been neglected in almost all SPGPE studies to date, often under the justification that it is expected to be weak in near-equilibrium scenarios~\cite{blakie_dynamics_2008,proukakis_finite-temperature_2008}. Only a handful of recent theoretical investigations have included energy damping, following developments in numerical techniques that have enabled simulation of the full SPGPE~\cite{rooney_numerical_2014,rooney_reservoir_2016,bradley_stochastic_2014,bradley_low-dimensional_2015, Mehdi:2021}. Notably, Ref.~\cite{rooney_reservoir_2016} studied the effect of both reservoir processes on the dissipation of a single vortex in a harmonically trapped three-dimensional gas, finding that energy damping was the dominant process in the regime studied. This is consistent with the results presented here, where we find that the energy-damping process is the dominant mechanism of point-vortex damping by two orders of magnitude. \par

\emph{Microscopic finite-temperature PVM.---}
Our model of finite-temperature point-vortex dynamics is constructed within the framework of SPGPE theory, explicitly including the energy-damping terms and neglecting the number-damping process - precisely the opposite approach taken in previous work. Our starting point is the quasi-2D SPGPE~\cite{bradley_low-dimensional_2015}, neglecting number-damping terms~\footnote{Although we neglect an explicit projector onto the low-energy subspace, since it does not significantly affect systems with homogeneous density, such projection is still implicitly included through the energy-cutoff dependence of \zm{the scattering kernel.}}:
\begin{align}
\label{eq:SPGPE_maintext} 
	i\hbar d\psi &= L[\psi] dt + \left(V_\varepsilon dt -\hbar dU_\varepsilon\right)\psi \,,
\end{align}
where $\psi(\textbf{x},t)$ is a classical field describing the finite-temperature superfluid dynamics in the plane $\mathbf{x}=(x,y)$, $L[\psi] = (-\hbar^2\nabla^2/(2m)+g|\psi|^2-\mu)\psi$ is the Gross-Pitaevskii (GP) operator, and spatial and temporal arguments are suppressed for brevity. $i\hbar d\psi = L[\psi] dt$ is precisely the GPE \zm{describing the conservative dynamics of the highly-occupied modes that host the vortices}, and remaining are reservoir terms. The energy-damping reservoir process leads to dissipation described by an effective scattering potential
\begin{align}
\label{eq:ED_potential}
	V_\varepsilon(\mathbf{x})= \hbar\int d^2\mathbf{x'} \varepsilon(\mathbf{x}-\mathbf{x'}) \: \frac{d\rho(\mathbf{x'})}{dt}
\end{align}
that damps changes in the fluid density $\rho = |\psi|^2$. Here $\varepsilon(\textbf{x})$ is the 2D scattering kernel. Associated with energy-damping dissipation is a Gaussian noise term representing incoherent thermal fluctuations, satisfying correlations $\langle dU_\varepsilon \rangle=0$ and $\hbar\langle dU_\varepsilon(\mathbf{x},t)dU_\varepsilon(\mathbf{x'},t') \rangle=2k_B T\varepsilon(\mathbf{x}-\mathbf{x'})\delta(t-t')dt$. Both the scattering potential and noise correlations are local in Fourier space, in which the kernel is $\tilde{\varepsilon}(\mathbf{k}) = 4 a_s^2 N_{\rm cut} e^{|l_z |\mathbf{k}|/2|^2}K_0(|l_z |\mathbf{k}|/2|^2)/\pi$, where $a_s$ is the s-wave scattering length, $N_{\rm cut}\equiv (e^{(\epsilon_{\rm cut}-\mu)/(k_B T)}-1)^{-1}$ is the number of reservoir atoms at the cutoff energy, $K_0$ is a modified Bessel function of the second kind, and $\sqrt{2}l_z$ is the $1/e$ transverse `thickness' of the atomic cloud~\cite{SupplementalMaterials}.
 
To derive a stochastic PV theory from Eq.~\eqref{eq:SPGPE_maintext}, we exploit the fact that the energy-damping terms form a stochastic potential, and hence can be added to the GP action as potential energy terms, \zm{giving} the Lagrangian density
\begin{align}
    \mathcal{L}dt &= \mathcal{L}_{\rm GP}dt + \hbar\left(V_\varepsilon dt-dU_\varepsilon \right)\rho \,, \label{lagrangian_density}
\end{align}
where $\mathcal{L}_{\rm GP}$ is the GP Lagrangian density. That is, the variational (least action) theory obtained by minimizing the action functional given by Lagrangian density Eq.~(\ref{lagrangian_density}) is precisely Eq.~\eqref{eq:SPGPE_maintext}. This allows for an analytic treatment of vortex dynamics, provided a suitable ansatz is chosen for the classical field $\psi$. We consider a system of isolated vortices \zm{on a homogeneous background,} each with a single quantum of circulation $\pm \hbar/m$, for which the variational theory of $\mathcal{L}_{\rm GP}$ reduces to the idealized PVM provided vortices are well separated~\cite{lucas_sound-induced_2014} \zm{-- a condition satisfied by modern 2D vortex experiments in homogeneous systems which operate within the point-vortex regime~\cite{moon_thermal_2015,johnstone_evolution_2019,gauthier_giant_2019,stockdale_universal_2020,kwon_sound_2021,reeves_turbulent_2022}}.

We assume $\rho(\mathbf{x},t) = \rho_0\left(1-\sum_{n} e^{-|\mathbf{x}-\mathbf{r}_n(t)|^2/2\xi^2}\right)$, where $\rho_0$ is the 2D background superfluid density and the healing length $\xi$ gives the scale of the vortex core~\cite{dalfovo_theory_1999}. This ansatz \zm{is entirely characterized by microscopic parameters and has a Gaussian vortex core, which is needed for analytic tractability and provides an excellent} approximation to the true vortex core in the region $|\mathbf{x}-\mathbf{r}_n|\lesssim \xi$ \zm{-- precisely where $d\rho/dt$, and hence dissipation described by Eq.~\eqref{eq:ED_potential}, is most significant.} 

Next, since the Fourier transform of $d\rho/dt$ is sharply peaked at the vortex core scale $|\textbf{k}| = \xi^{-1}$ for this density ansatz, we can treat the scattering kernel as approximately constant in Fourier space: $\tilde{\varepsilon}(\mathbf{k})\approx \tilde{\varepsilon}(\xi^{-1})$. Consequently, $V_\varepsilon \approx 2 N_{\rm cut}\hbar \sigma_{\rm ED} d\rho/dt$, where $\sigma_{\rm ED} \equiv  \sigma_s e^{l_z^2/(2\xi)^2}K_0(l_z^2/(2\xi)^2)/(2\pi)$ is an effective \zm{2D} scattering cross section for the energy-damping process in terms of the s-wave scattering cross section $\sigma_s=8\pi a_s^2$~\cite{Pethick2008}. A similar argument justifies treating the energy-damping noise \zm{correlator} as approximately local in space~\cite{SupplementalMaterials}: $\hbar\langle dU_\varepsilon(\mathbf{x},t)dU_\varepsilon(\mathbf{x'},t) \rangle\approx4k_B T\sigma_{\rm ED} N_{\rm cut}\delta(\mathbf{x}-\mathbf{x'})dt$. 
 
Under the above approximations, the spatial integrals over the damping and noise terms can be computed analytically. In the point-vortex limit of well-separated vortices $r_{nm}^2\equiv|\mathbf{r}_n-\mathbf{r}_m|^2\gg\xi^2$, the least action theory of the resulting Lagrangian gives the following stochastic point-vortex equation up to corrective factors of order $\mathcal{O}(\alpha_\varepsilon^2,e^{-r_{nm}^2/4\xi^2})\ll 1$~\cite{SupplementalMaterials}:
\begin{align}
	\label{eq:stochasticdPV_withED}
	d\mathbf{r}_n &= \left(\mathbf{v}^0_n - \alpha_\varepsilon q_n \mathbf{\hat{z}}{\times} \mathbf{v}^0_n\right) dt+\sqrt{2\eta}d\mathbf{w}_n,
\end{align}
which includes a vortex damping term precisely of the form in Eq.~\eqref{eq:dPVModel_Old}, with mutual friction coefficient $\alpha_\varepsilon \equiv \sigma_{\rm ED} \rho_0 N_{\rm cut}/2$.
The stochastic point-vortex equation also includes a diffusive noise process $d\mathbf{w}_n=(dW_n^x,dW_n^y,0)$, driven by Gaussian random variables with zero mean and correlations given by $\langle dW_n^\alpha(t)dW^\beta_m(t')\rangle =\delta_{\alpha\beta}\delta_{nm}\delta(t-t')dt$. The diffusion coefficient $\eta \equiv \alpha_\varepsilon k_B T/(2\pi\hbar\rho_0)$ describes Brownian motion of the vortices due to thermal fluctuations; \zm{further approximating the background flow with its average over the vortices}, the convective position variance of each vortex grows diffusively: $\langle \Delta \textbf{r}_n^2 \rangle = 4\eta t$~\cite{SupplementalMaterials}.   \par 

The microscopic expression for the mutual friction coefficient has a simple physical interpretation, as it is proportional to the product of the per-particle probability of energy-damping scattering $\sigma_{\rm ED}\rho_0$ and the number of thermal reservoir atoms at the cutoff $N_{\rm cut}$. That is, the mutual friction coefficient is proportional to the rate of number-conserving two-body scattering events between the atoms in the condensed modes and atoms in the reservoir. It scales both with the number of reservoir atoms and condensed atoms, and recovers the idealized (non-dissipative) PVM as $T\rightarrow 0$. \par                                   
\begin{figure}[!t]
	\centering
   \includegraphics[width=0.99\columnwidth]{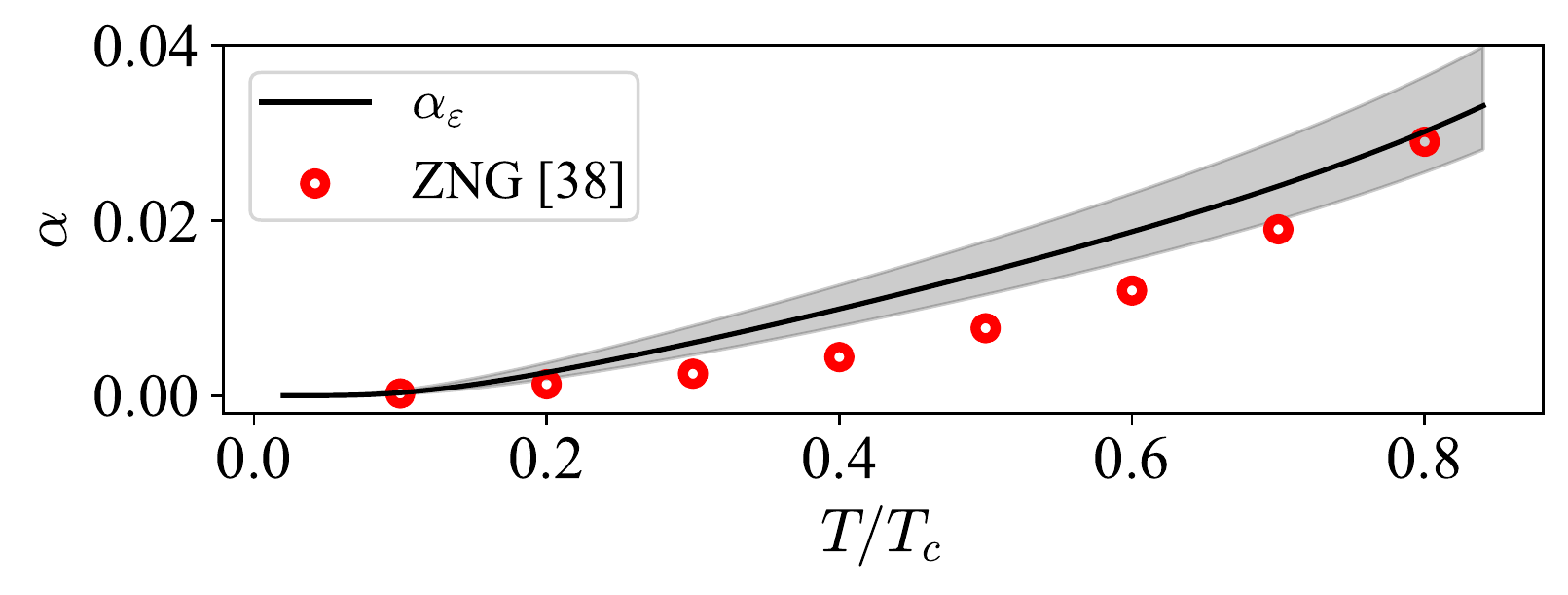}
	\caption{Comparison of microscopic mutual friction $\alpha_\varepsilon$ (estimated with $\epsilon_{\rm cut}=2\mu$) to mutual friction determined phenomenologically by fitting the damped PVM Eq.~(\ref{eq:dPVModel_Old}) to dynamical ZNG simulations \cite{jackson_finite-temperature_2009}. We observe good quantitative agreement, with slight disagreement at intermediate temperatures. The microscopic prediction of $\alpha$ varies slightly with the choice of energy cutoff; the shaded region denotes a $15\%$ variation of $\epsilon_{\rm cut}$.}
   \label{fig:JacksonComparison}
\end{figure}

Additionally, the effective energy-damping cross section $\sigma_{\rm ED}$ is a monotonically decreasing function of the transverse thickness $l_z$, and thus so too is the mutual friction coefficient $\alpha_\varepsilon$. Although the validity of the quasi-2D SPGPE requires $l_z\lesssim \xi$, we find our stochastic PV theory can quantitatively capture vortex damping in the much less restrictive regime $l_z \lesssim 10\xi$, consistent with previous work~\cite{rooney_suppression_2011}. We demonstrate this in Fig.~\ref{fig:JacksonComparison} by comparing our microscopic expression for the mutual friction coefficient to values phenomenologically extracted from numerical Zaremba-Nikuni-Griffin (ZNG) kinetic theory simulations via a fit to Eq.~(\ref{eq:dPVModel_Old})~\cite{jackson_finite-temperature_2009}. There is excellent quantitative agreement with the fitted ZNG values of $\alpha$, despite the system considered in Ref.~\cite{jackson_finite-temperature_2009} having a significant three-dimensional extent, $l_z{\sim}10\xi$. Figure~\ref{fig:JacksonComparison} also demonstrates that the value of $\alpha_\varepsilon$ does not strongly depend on the precise choice of $\epsilon_{\rm cut}$~\cite{SupplementalMaterials}, varying weakly as $\epsilon_{\rm cut}$ is varied by $15\%$. 
 
Although the SPGPE reservoir dissipation and noise terms satisfy the fluctuation-dissipation relation, this same relation does not hold for the damping and noise terms in the stochastic PV equation Eq.~\eqref{eq:stochasticdPV_withED}, as the thermal equilibrium of the atomic cloud corresponds to a system with no vortices (for a non-rotating thermal cloud). As noted in Ref.~\cite{reeves_turbulent_2022}, which used a stochastic PVM with experimentally-fitted coefficients, the noise term in Eq.~\eqref{eq:stochasticdPV_withED} is dissipative and can be understood as an effective viscosity term.

Historically, a stochastic PVM of similar form to Eq.~(\ref{eq:stochasticdPV_withED}) was considered within the context of superfluid helium \cite{Ambegaokar1980}. However, the inclusion of stochastic noise in Ref.~\cite{Ambegaokar1980} was not derived from microscopic theory, but motivated by fluctuation-dissipation arguments to generalize the phenomenological Hall-Vinen-Iordanskii equations~\cite{hall_rotation_1956,vinen_mutual_1957,iordanskii_mutual_1966}. Tractable microscopic models are lacking for superfluid helium, necessitating a phenomenological treatment of the damping that can only qualitatively describe experiment~\cite{barenghi_friction_1983}.

\emph{Experimental comparison.---}
\begin{figure}[!t]
	\centering
   \includegraphics[width=\columnwidth]{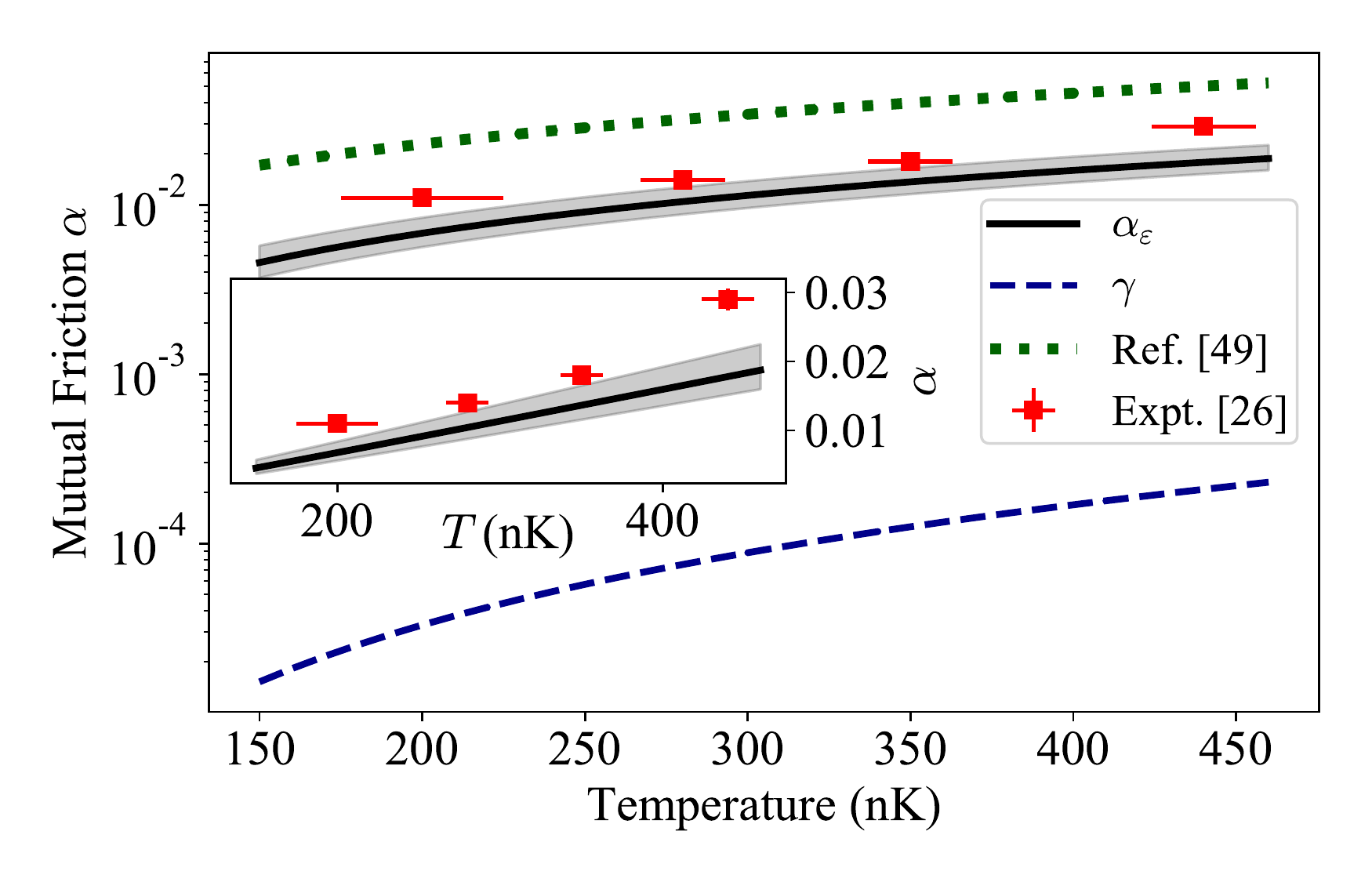}
	\caption{Microscopic mutual friction contributions of number damping $\alpha_\gamma=\gamma$ and energy damping $\alpha_\varepsilon$ (both estimated with $\epsilon_{\rm cut}=2\mu$) compared to experimental measurements reported in Ref.~\cite{moon_thermal_2015}. Shaded region gives variation of $\alpha_\varepsilon$ for $15\%$ change in $\epsilon_{\rm cut}$. The number-damping estimate is several orders of magnitude smaller than the experimentally-measured values; much closer agreement is given by the energy-damping estimate. The inset more clearly compares the experimental results to the energy-damping prediction, demonstrating agreement within $\approx 20{-}40\%$ of the experimentally-measured values. For comparison, the analytical estimate of Ref.~\cite{fedichev_dissipative_1999} disagrees with experiment by approximately a factor of two.}
   \label{fig:ExpComparisonRates}
\end{figure}
Reference~\cite{moon_thermal_2015} reports experimental measurements of the mutual friction coefficient, obtained by fitting experimentally-measured trajectories of a pair of like-sign vortices in a harmonic trap to the predictions of Eq.~\eqref{eq:dPVModel_Old}. 
Figure~\ref{fig:ExpComparisonRates} compares these experimentally-measured values to the analytic prediction of our microscopic model $\alpha_\varepsilon=\sigma_{\rm ED} \rho_0 N_{\rm cut}/2$. For the temperature range considered in the experiment, the noise term in Eq.~\eqref{eq:stochasticdPV_withED} is negligible (\zm{$\eta\sim 10^{-5}\hbar/m$}), despite strong damping.

Our microscopic prediction is within $\sim20{-}40\%$ of the experimentally-measured values; a remarkable result given the experimental measurements were affected by density inhomogeneity and temperature-dependent vortex precession due to the harmonic trapping~\footnote{\zm{Although corrections due to weak density gradients are small, they could be included under the local density approximation, giving spatially-dependent damping and diffusion coefficients based on local density at each vortex.}}. 
In comparison, the mutual friction coefficient predicted from number-damping reservoir interactions ($\alpha_\gamma=\gamma$) is several orders of magnitude smaller than the experimentally-determined values, across the entire temperature range. This justifies the neglect of the number-damping process in our stochastic PV theory Eq.~(\ref{eq:stochasticdPV_withED}). \zm{We may understand this result by noting the number-damping reservoir process drives equilibration against imbalance between the chemical potential of the low-energy atomic modes and the reservoir, which is significant during condensate growth. For vortex excitations in an otherwise equilibrated gas, which have a very low effective chemical potential, damping is therefore dominated by the number-conserving reservoir process which directly opposes changes in the atomic density.  }
  \par 

Our microscopic prediction $\alpha_\varepsilon$ also gives closer quantitative agreement than the theoretical estimate of $\alpha \approx (n_{\rm th }/n_{0})\sqrt{\mu/(k_BT)}$ by Ref.~\cite{fedichev_dissipative_1999}, where $n_{\rm th }/n_{0}$ is the ratio of the three-dimensional densities of the thermal cloud and condensate, respectively. This estimate is derived from low-energy perturbation theory and can only be phenomenologically extended to compare to the experiment of Ref.~\cite{moon_thermal_2015}, where a significant thermal fraction was present.  \par 

The close quantitative agreement between the microscopic prediction $\alpha_\varepsilon=\sigma_{\rm ED} \rho_0 N_{\rm cut}/2$ and this experiment validates the dissipative component of our stochastic PV equation, confirming that energy-damping reservoir interactions are the underlying microscopic mechanism to mutual friction in 2D atomic superfluids. It is therefore unsurprising that previous attempts to estimate mutual friction rates from first principles, which have generally ignored energy-damping interactions, have not been successful.

Consequently, previous microscopic modelling of thermal friction in 2D quantum vortices that neglected energy damping must be treated as phenomenological. It will therefore be important to revisit previous microscopic 2DQT modelling, such as in Ref.~\cite{groszek_decaying_2020}, and make quantitative predictions with the explicit inclusion of energy-damping interactions. \par

\zm{ 
\emph{Vortex diffusivity at the BKT transition.---} At higher temperatures near the transition to superfluidity, vortex diffusion driven by critical thermal fluctuations is expected to play a more significant role. In particular, the diffusion coefficient $\eta$ is an essential parameter in dynamic corrections to static BKT theory~\cite{Ambegaokar1980,Adams1987,Gillis1989}, yet it is currently treated as a phenomenological fitting parameter in the context of ultracold atomic gases~\cite{Wu2020}. Our work provides a microscopic expression for $\eta$ and sets a theoretical foundation for future experiments probing the BKT regime. For example, we can predict the value of $\eta$ for a weakly-interacting Bose gas at the BKT transition temperature $T_c^{\rm BKT}=2\pi\rho_0\hbar^2/(m k_B \ln(360/\tilde{g}))$~\cite{Prokofev2001}, where $\tilde{g}=\sqrt{8\pi}a_s/l_z$.  
For typical experimental parameters (e.g. Ref.~\cite{Ville2018}), our microscopic expression predicts $\eta\sim10^{-2}\hbar/m$ at the BKT transition -- two orders of magnitude smaller than in strongly-interacting superfluid helium films~\cite{Ambegaokar1980,Adams1987}.}

\emph{Conclusions and Outlook.---}
We have provided a microscopic foundation for mutual friction and thermal diffusion of 2D vortex motion in atomic superfluids, based on often-neglected finite-temperature interactions with a static thermal reservoir. We derived a stochastic point-vortex theory, which gives an analytic expression for the mutual friction coefficient that compares excellently with available experimental data. The damped evolution in this theory is consistent with previous phenomenological modelling, validating the mutual friction concept in studies of turbulent atomic superfluids. These results profoundly impact future theory of vortex dynamics in atomic superfluids, \zm{showing the importance of energy-damping interactions for quantitative  understanding of dissipation. Crucially, our microscopic theory allows experimentally-testable predictions of 2D quantum vortex dynamics without any fitted parameters.}\par 

2DQT experiments have evolved significantly and now routinely study 2D vortex dynamics in homogeneous systems \cite{johnstone_evolution_2019,gauthier_giant_2019,stockdale_universal_2020,kwon_sound_2021,reeves_turbulent_2022}, allowing further tests of our theory. \zm{In particular, experimentally measuring the diffusion coefficient $\eta$ is an important test of stochastic point-vortex theory}. In principle, $\eta$ can be extracted from the dynamics of a co-rotating pair of same-sign vortices by tracking the drift in the center-of-mass position of the pair. The thermal noise identified sets a floor for the total noise that will include other sources, such as technical noise in the trapping potential and incoherent density fluctuations. Observing the fundamental point-vortex noise poses an interesting challenge for future experimental study, \zm{and an important test of BKT physics in ultracold gases}.

A deeper understanding of thermal friction's role in quantum turbulence could require further theoretical investigations into the effect of reservoir interactions on vortex dynamics outside of the point-vortex regime, 
for which a numerical approach will probably be required. \zm{Beyond its influence on vortex core motion, energy damping will also couple strongly to other compressible excitations~\cite{mcDonald_brownian_2016,mcdonald_dynamics_2020}, with important implications for weak-wave turbulence in quantum fluids~\cite{nazarenko_wave_2006,GalkaEmergence2022_Preprint}}. \par

\emph{Acknowledgements.---}
We acknowledge insightful discussions with John Close, Tyler Neely, and Matthew Reeves. This research was undertaken with the assistance of resources and services from the National Computational Infrastructure (NCI), which is supported by the Australian Government. ZM is supported by an Australian Government Research Training Program (RTP) Scholarship. ASB acknowledges financial support from the Marsden Fund (Grant No. UOO1726) and the Dodd-Walls Centre for Photonic and Quantum Technologies. SSS is supported by an Australian Research Council Discovery Early Career Researcher Award (DECRA), Project No. DE200100495.

\bibliography{bib}


\pagebreak
\renewcommand{\thefigure}{[S\arabic{figure}]} 
\renewcommand{\theequation}{S\arabic{equation}} 
\setcounter{equation}{0}
\setcounter{figure}{0}
\widetext
\pagebreak
   \begin{center}
      \large{\bf Supplemental Material: Mutual friction and diffusion of two-dimensional quantum vortices}\\
   \end{center}
   \vspace*{2.0pt}

In this supplemental material we provide (1) a brief review of the full three-dimensional stochastic Gross-Pitaevskii equation (SPGPE) that includes both number-damping and energy-damping reservoir mechanisms, (2) a brief review of the quasi-2D SPGPE (Eq.~(2) of the main text) and approximate treatment of the scattering kernel, (3) the derivation and numerical validation of the stochastic point-vortex equation (PVM) (Eq.~(5) of the main text), (4) a proof showing that vortex evolution under the stochastic point-vortex equation corresponds to Brownian motion in the mean-field limit, and (5) estimation of atomic cloud parameters for calculations of the mutual friction coefficient.

\section{(1) Stochastic projected Gross-Pitaevskii (SPGPE) Theory}
The SPGPE is a first-principles reservoir theory that quantitatively describes a finite-temperature ultracold Bose gas~[24].
Within this framework, highly-populated modes of the quantum field (generally $\gtrsim 1$ atoms on average) are treated as a coherent classical field $\psi$, which interacts with an incoherent thermal reservoir composed of the remaining sparsely-occupied high-energy modes. This leads to a stochastic equation of motion for the classical field $\psi$, which in Stratonovich form is
\begin{align}
\label{eq:SPGPEFull}
    i\hbar d\psi &= \mathcal{P}\big{\{} (1-i\gamma)(\mathcal{L}-\mu)\psi dt + i\hbar d\xi_\gamma (\textbf{r},t) + V_\varepsilon(\textbf{r},t)\psi dt - \hbar\psi dU_{\varepsilon}(\textbf{r},t) \big{\}} \,,
\end{align}
where $\mathcal{L}=H_0+g|\psi|^2$ for the single-particle Hamiltonian $H_0$. Here $g=4\pi a_s \hbar^2/m$ is the two-body interaction strength for an s-wave scattering length $a_s$. The explicit inclusion of the projector $\mathcal{P}$ ensures the dynamic separation of the field into a low-energy coherent region and incoherent reservoir. The noise terms $d\xi_\gamma$ and $dU_{\varepsilon}$ correspond to incoherent thermal fluctuations from the number-damping and energy-damping processes, respectively, and coupled with the deterministic dissipation terms ultimately drive any initial state to a steady-state at thermal equilibrium within the grand-canonical ensemble at temperature $T$ and chemical potential $\mu$. The strengths of the number-damping and energy-damping dissipation processes are characterized by the dimensionless quantity $\gamma$ and the length-squared quantity $\mathcal{M}$, respectively. These parameters can be \emph{a priori} determined from the reservoir chemical potential $\mu$, temperature $T$, and the energy cutoff $\epsilon_\text{cut}$~[31]:
\begin{align}
\label{eq:ND_Rate}
    \gamma&=\frac{8a_s^2}{\lambda_\text{dB}^2}\sum_{j=1}^\infty \frac{e^{\beta \mu (j+1)}}{e^{2\beta\epsilon_\text{cut}j}} \Phi[e^{\beta (\mu-2\epsilon_\text{cut})},1,j] \,, \\
    \label{eq:ED_Rate}
    \mathcal{M} &= \frac{16\pi a_s^2}{\exp\left(\frac{\epsilon_{\rm cut}-\mu}{k_B T}\right)-1} \,,
\end{align}
where $a_s$ is the s-wave scattering length, $\beta=1/(k_\text{B} T)$, $\lambda_\text{dB}=\sqrt{2\pi\hbar^2/(m k_\text{B} T)}$ is the thermal de~Broglie wavelength, and $\Phi[z,x,a]$ is the Lerch transcendent.

In the main text, we define $N_{\rm cut}\equiv(e^{(\epsilon_{\rm cut}-\mu)/(k_B T)}-1)^{-1}$ as the number of reservoir atoms at the cutoff energy, as it is the thermal equilibrium number distribution for an ideal gas evaluated at the cutoff energy. This is a good estimate of the true number of atoms at the cutoff energy, as the cutoff energy $\epsilon_{\rm cut}$ should be sufficiently large compared to $\mu$ such that high-energy modes very close to the cutoff are essentially non-interacting (see the discussion below). The energy-damping coefficient can then be written as $\mathcal{M}=2\sigma_sN_{\rm cut}$, where $\sigma_s\equiv 8\pi a_s^2$ is s-wave scattering cross section.

The energy-damping dissipation process is described by an effective potential term $V_\varepsilon$: 
\begin{equation}
    \label{scattering potential}
    V_\varepsilon(\textbf{r},t) = -\hbar\int d^3 \textbf{r}' \varepsilon_\textrm{3D}(\textbf{r}-\textbf{r}') \nabla_{\textbf{r}'}\cdot \textbf{j}(\textbf{r}',t)\,,
\end{equation}
which is a convolution between the divergence of the particle current
\begin{equation}
    \textbf{j}(\textbf{r},t) = \frac{i\hbar}{2m}[\psi\nabla\psi^*-\psi^*\nabla\psi]
\end{equation}
and the scattering kernel 
\begin{equation}
    \varepsilon_\textrm{3D}(\textbf{r}) = \frac{\mathcal{M} }{(2\pi)^3}\int d^3 \textbf{k} \frac{e^{i \textbf{k}\cdot \textbf{r}}}{|\textbf{k}|} \,.
\end{equation}
The noise terms in the SPGPE are random Gaussian variables with zero mean and correlations:
\begin{align}
\langle d\xi_\gamma^*(\mathbf{r},t)d\xi_\gamma(\mathbf{r}',t')\rangle &= \frac{2\gamma k_\text{B} T}{\hbar} \delta (\mathbf{r}-\mathbf{r'}) \delta(t-t')dt \,, \\ \label{eq:ED_NoiseCorr}
    \langle dU_{\varepsilon}(\mathbf{r},t)dU_{\varepsilon}(\mathbf{r'},t')\rangle &= \frac{2k_\text{B}T}{\hbar}\varepsilon_\textrm{3D}(\mathbf{r}-\mathbf{r'})\delta(t-t')dt \,.
\end{align}
Note that the number-damping noise $d\xi_\gamma$ is complex, whereas the energy-damping noise $dU_{\varepsilon}$ is real-valued.

To zeroth order in the reservoir processes, the SPGPE satisfies the continuity equation
\begin{align}
	\nabla \cdot \mathbf{j} + \frac{\partial\rho}{\partial t} =0 \,,
\end{align}
where $\rho=|\psi|^2$ is the fluid density (the leading correction occurs at order $\gamma \ll 1$). Therefore, at leading order in the dissipation parameters, the energy-damping potential directly opposes changes in the density:
\begin{equation}
    \label{eq:approx_continuity}
    V_\varepsilon(\textbf{r},t)dt = \hbar\int d^3 \textbf{r}' \varepsilon(\textbf{r}-\textbf{r}') d\rho(\textbf{r}',t) + \mathcal{O}(\gamma \mathcal{M}) \,.
\end{equation}
This relation is used below in our derivation of the stochastic point-vortex equation.

\subsection{Determination of the energy cutoff}
Determining the value of the energy cutoff $\epsilon_{\rm cut}$ for a particular experiment is a non-trivial yet essential aspect of first-principles modelling with the SPGPE. To ensure the validity of the SPGPE framework, the choice of cutoff must satisfy two key properties. Firstly, $\epsilon_{\rm cut}$ must be chosen such that each of the modes in the low-energy region are appreciably occupied, i.e. have occupation no fewer than $\mathcal{O}(1)$ atoms~[24]. 
Secondly, the cutoff should be sufficiently large (compared to $\mu$) to ensure the interacting modes of the system are contained within the low-energy coherent region. The latter requirement is typically satisfied for $\epsilon_{\rm cut}\gtrsim 2\mu$~[29].

These constraints do not uniquely specify a particular energy cutoff, but rather tightly constrains appropriate choices of $\epsilon_{\rm cut}$. In principle, this means calculations in the SPGPE framework will depend weakly on the precise choice of $\epsilon_{\rm cut}$. In practise, it is therefore important for first-principles SPGPE calculations to demonstrate robustness of results to small variations of $\epsilon_{\rm cut}$ (on the order of $10\%$) -- see, for example, Refs.[5,24,31,42]
. In the main text results are shown for a $15\%$ variation in $\epsilon_\text{cut}$. 

For the comparison to Ref.~[26] 
presented in Fig.~3 of the main text, we find the choice of $\epsilon_\text{cut}=2\mu$ to satisfy the two requirements described above, across the temperature range considered.
Significantly increasing the cutoff beyond this value results in the highest-energy modes becoming too sparsely occupied, particularly for the lower range of temperatures considered. For example, setting $\epsilon_\text{cut}=3\mu$ results in $N_{\rm cut}\approx 0.4$ for the $T=200$nK in Fig.~3 of the main text. Significantly reducing the cutoff below $2\mu$ will result in a number of appreciably-occupied interacting modes of the system inappropriately becoming part of the incoherent region.

\section{(2) Quasi-2D SPGPE and approximate energy-damping potential}
\label{append:Approx2DKernel}
For studies of two-dimensional systems, it is convenient to work with a quasi-2D form of the SPGPE, where the transverse ($z$) degrees of freedom are integrated out. The resulting quasi-2D SPGPE (Eq.~(2) of the main text) has the same form as the 3D SPGPE Eq.~(\ref{eq:SPGPEFull}) with the following modified 2D scattering kernel~[41]:
\begin{equation}
    \varepsilon(\textbf{x}) = \frac{1}{2\pi}\int d^2 \textbf{k} \, e^{i \textbf{k}\cdot \textbf{x}}\underbrace{\left[\frac{\mathcal{M}}{(2\pi)^2}F\left(\frac{(l_z |\textbf{k}|)^2}{4}\right)\right]}_{\tilde \varepsilon(\textbf{k})} \,,
\end{equation}
where $F(x)=e^xK_0(x)$ with $K_0$ a modified Bessel function of the second kind. This reduction to 2D assumes the transverse field can be described by a Gaussian of $1\sigma$ radius $l_z$, as described in Section 5 of this document.

For the study of 2D vortex dynamics, we only require that the lengthscale be of the order of the healing length $l_z\approx\xi$, which ensures that Kelvin waves along the vortex filaments are suppressed~[48].
Since this is a much less restrictive condition than the oblate confinement need to realize a thermodynamically two-dimensional gas (i.e. the BKT transition), experiments can investigate two-dimensional vortex dynamics in a convenient regime where condensate fraction, temperature, etc. are all well defined~[32].\par 

The convolution with the scattering kernel in the energy-damping potential adds a level of complexity that prevents most integrals involving $V_\varepsilon$ to be analytically solved. In the main text we treat this by approximating the kernel as flat in Fourier space, evaluated at the vortex core scale $k=\xi^{-1}$. This results in a simplified form of the kernel:
\begin{align}
\label{eq:approximate_edkernel}
	\varepsilon(\mathbf{x}) &\approx 2\pi \tilde{\varepsilon}(\xi^{-1})  \delta(\mathbf{x}) \,, \\
	&= 2\sigma_{\rm ED} N_{\rm cut}\delta(\mathbf{x}) \,,
\end{align}
where $\sigma_{\rm ED}=\sigma_s F\left(\frac{l_z^2}{4\xi^2} \right)/(2\pi)$ is the effective energy-damping scattering cross section defined in the main text, and the factor of $2\pi$ in the first line arises due to the convolution theorem for the two-dimensional Fourier transform (in the unitary transform convention).

\section{(3) Full derivation of stochastic point-vortex equation}
\begin{figure}
    \centering
    \includegraphics[width=0.9\textwidth]{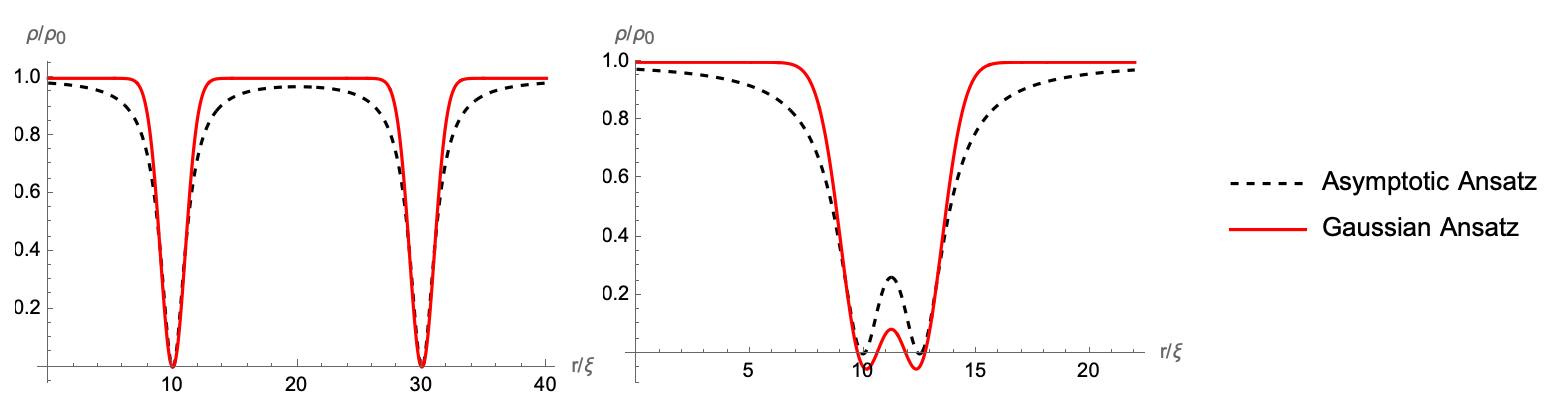}
    \caption{Comparison of the Gaussian density ansatz Eq.~\eqref{eq:DensityAnsatz} (red) to the analytic asymptotic solution $\rho(\bm{x}) \approx \rho_0\prod_n |\bm{x}-\bm{r}_n|^2/(|\bm{x}-\bm{r}_n|^2+0.82^{-2})$~[32] (black, dashed), for a pair of well-separated vortices (left) and vortices separated by $2.5\xi$ (right). For the former case, the ansatz is an excellent description of density within $1\xi$ of the vortex cores and is positive definite. In the latter case, the density becomes slightly negative as there is significant overlap between the two vortex cores.}
    \label{fig:Ansatz}
\end{figure}

Our derivation of the stochastic point-vortex equation described in the main text uses an effective Lagrangian formulation of the quasi-2D SPGPE and the approximate form of the energy-damping kernel Eq.~(\ref{eq:approximate_edkernel}) described in the previous section. Since idealized point-vortex dynamics can be rigorously derived from the GPE in the limit of well-separated vortices~[44],
we describe the non-dissipative Gross-Pitaevskii dynamics of our system via the point-vortex Lagrangian~\footnote{This expression has an additional factor of two compared to that of Ref.~[44], 
which is required for agreement between the conserved energy of the above Lagrangian and the GPE Hamiltonian. We have numerically confirmed this for the case of a well-separated vortex dipole.}:
\begin{align}
    L_\text{PV} &= 2\pi\hbar\rho_0 \bigg( \sum_n \frac{q_n}{2}\epsilon_{ij} \dot{X}_n^i X_n^j + \frac{\hbar}{m}\sum_{m\neq n}q_n q_m \log\frac{|\mathbf{r}_m-\mathbf{r}_n|}{l} \bigg),
\end{align}
where $\epsilon_{ij}$ is the Levi-Civita symbol, $\rho_0$ is the background 2D density of the fluid, and  $\mathbf{r}_n=(X_n,Y_n)^T$ and $q_n=\pm 1$ is the 2D position vector and charge of the $n$-th vortex, respectively. The box size $l$ is included here as a cutoff to regularize the PV theory; it does not explicitly appear at any point in our derivation below. The SPGPE energy-damping terms are real-valued and can be treated as potential energy terms that can simply be added to $L_\text{PV}$, giving the effective point-vortex Lagrangian
\begin{align}
    L_\text{eff} &= L_\text{PV} +  L_\text{damping} + L_\text{noise}
\end{align}
where the additional reservoir terms due to energy damping are:
\begin{align}
    L_\text{damping} &\equiv \int d^2 \mathbf{x} \, V_\varepsilon(\mathbf{x},t)\rho(\mathbf{x},t) \label{eq:Ldamping} \\
    L_\text{noise} dt &\equiv -\hbar \int d^2\mathbf{x} \, dU_{\varepsilon}(\mathbf{x},t) \rho(\mathbf{x},t) \,.
\end{align}
These terms do not depend upon the full classical field $\psi$ and can therefore be solved with an appropriate ansatz for the 2D fluid density $\rho=|\psi|^2$ alone. We choose a Gaussian ansatz for $\rho$ that separates the contribution of the vortices from the infinite background:
\begin{align}
\label{eq:DensityAnsatz}
	\rho(\textbf{x}) = \rho_0\left[1-\sum_{n} \exp\left(-\frac{|\mathbf{x}-\mathbf{r}_n(t)|^2}{2\xi^2}\right)\right],
\end{align}
which for a single vortex agrees well with the exact GPE solution for the core in the range $|\mathbf{x}-\mathbf{r}_n(t)| \lesssim \xi$. Although this ansatz is not strictly non-negative, \zm{it is well approximated as such} in the point-vortex regime where vortices are well separated (which is required for the validity of the point-vortex model in general) \zm{as illustrated in Fig.~\ref{fig:Ansatz}}. As we show below, this ansatz allows the above integrals to be solved \emph{exactly} for an $N$-vortex system. An additional benefit of this form of ansatz is that the infinite background term does not need to be manually discarded, as only derivatives of the density contribute to the final equation of motion.

We first focus on the damping term Eq.~(\ref{eq:Ldamping}), which for the above ansatz and $\varepsilon(\mathbf{x}) \approx 2\sigma_{\rm ED} N_{\rm cut}\delta(\mathbf{x})$ is given by:
\begin{align}
    L_\text{damping} &= 2\alpha_\varepsilon\pi\hbar\rho_0\sum_{nm}\exp\left(-\frac{r_{mn}^2}{4\xi^2}\right) \left(\dot{X}_m\delta x_{mn}+\delta y_{mn}\dot{Y}_m \right),
\end{align}
where we are using the shorthand $\delta x_{mn}\equiv X_m-X_n$, $\delta y_{mn}\equiv Y_m-Y_n$, $r_{mn}^2 \equiv \delta x_{mn}^2 + \delta y_{mn}^2$, and identified the expression for the mutual friction coefficient $\alpha_\varepsilon\equiv \sigma_{\rm ED}N_{\rm cut}\rho_0/2 = \sigma_s N_{\rm cut}\rho_0F(l_z^2/(4\xi^2))/(4\pi)$ defined in the main text. Neglecting $L_\text{noise}$ for now, we can derive the dissipative dynamics by taking the Euler-Lagrange equations with respect to $L_\text{PV}+L_\text{damping}$:
\begin{align}
\label{eq:ELEqns_Ideal} 
 q_n\bigg[\begin{pmatrix}
	 -\dot{Y}_n\\ 
	 \dot{X}_n
	\end{pmatrix} 
	+\frac{\hbar}{m}\sum_{m \neq n} \frac{q_m}{r_{mn}^2}
	\begin{pmatrix}
	\delta x_{nm}\\ 
	\delta y_{nm}
	\end{pmatrix} \bigg] =\frac{\alpha_\varepsilon}{2\xi^2}\sum_{m}\exp\left(-\frac{r_{mn}^2}{4\xi^2}\right)\begin{pmatrix}
	  \dot{X}_m(2\xi^2-\delta x_{mn}^2)-\dot{Y}_m\delta x_{mn}\delta y_{mn} \\ 
	 \dot{Y}_m(2\xi^2-\delta y_{mn}^2)-\dot{X}_m \delta x_{mn}\delta y_{mn}
	\end{pmatrix},
\end{align}
where we have cancelled common factors of $2\pi\hbar\rho_0$. In the point-vortex limit $r_{mn}^2 \gg \xi^2$, we may make the approximation $\exp[-r_{mn}^2 / (4 \xi^2)] \approx \delta_{mn}$, resulting in a very simple set of equations:
\begin{align}
	\begin{pmatrix}
	 \dot{X}_n\\ 
	 \dot{Y}_n
	\end{pmatrix} 
	\approx \frac{\hbar}{m}\sum_{m \neq n}\frac{q_m}{r_{mn}^2}
	\begin{pmatrix}
	 -\delta y_{nm}\\ 
	\delta x_{nm}
	\end{pmatrix}
	+\frac{\alpha_\varepsilon}{q_n}\begin{pmatrix}
	 \dot{Y}_n\\ 
	 -\dot{X}_n
	\end{pmatrix} \,.
\end{align}
Since the first term on the right-hand side is exactly the background superfluid velocity at the $i$-th vortex $\mathbf{v}_i^0$, we may write this equation as (assuming $q_n=\pm 1$):
\begin{align}
	\mathbf{\dot{r}}_n &= \mathbf{v}^0_n -\alpha_\varepsilon q_n \mathbf{\hat{z}}{\times}\dot{\mathbf{r}}_n  \notag \\
	    \label{eq:dissipative-point-vortex}
	&= \mathbf{v}^0_n -\alpha_\varepsilon q_n\mathbf{\hat{z}}{\times}\mathbf{v}^0_n + \mathcal{O}\left( \alpha_\varepsilon^2\right) \,,
\end{align}
where in the last line we substituted in the zeroth-order result $\mathbf{\dot{r}}_n=\mathbf{v}^0_n+\mathcal{O}(\alpha_\varepsilon)$. This equation is precisely the damped PVM (Eq.~(1) of the main text), with mutual friction coefficient $\alpha_\varepsilon$.

We now derive the full effective Lagrangian by first turning our attention to the energy-damping noise term. In contrast to the approach for the damping term, we will find it convenient to evaluate the spatial integrals after first taking the Euler-Lagrange equations with respect to the full effective Lagrangian $L_{\rm eff}$, which gives the equations of motion:
\begin{align}
	\begin{pmatrix}
	 dX_n\\ 
	 dY_n
	\end{pmatrix} 
	\approx \underbrace{\frac{\hbar}{m}\sum_{m \neq n}\frac{q_m}{r_{nm}^2}
	\begin{pmatrix}
	 -\delta y_{nm}\\ 
	\delta x_{nm}
	\end{pmatrix}dt
	+\alpha_\varepsilon q_n\begin{pmatrix}
	  dY_n\\ 
	 -dX_n
	\end{pmatrix}}_{\text{RHS of Eq.~(\ref{eq:dissipative-point-vortex})}} + d\mathbf{U}_n(t)\,,
\end{align}
where we have again assumed $q_n=\pm 1$, and defined the stochastic noise vector
\begin{align}
    d\mathbf{U}_n(t) &\equiv \frac{1}{2\pi \rho_0 q_n} \int d^2\mathbf{x} \, dU_{\varepsilon}(\mathbf{x},t) \begin{pmatrix}
	 -(\partial \rho(\mathbf{x}) / \partial Y_n) \\ 
	(\partial \rho(\mathbf{x})  / \partial X_n) 
	\end{pmatrix}.
\end{align}
We will now consider the noise correlations of this stochastic noise vector using Eq.~\eqref{eq:ED_NoiseCorr}. To simplify our expressions, we denote the $i$th element of the vectors $d\mathbf{U}_n(t)$ and $\textbf{r}_n = (X_n, Y_n)^T$ by $dU^i_n(t)$ and $X_n^i$, respectively, and define $\sigma^{ij}=1$ if $i = j$ and $\sigma^{ij} = -1$ if $i \neq j$. 

Following from the properties of $dU_\varepsilon$, $d\mathbf{U}_n(t)$ is a Gaussian noise vector with zero mean and correlations:
\begin{subequations}
\begin{align}
\label{eq:projected_noisecorr}
    \langle dU^i_n(t) dU^j_m(t') \rangle &= \frac{\sigma^{ij}}{(2\pi\rho_0)^2} \int d^2\mathbf{x} \int d^2\mathbf{y} \, \frac{\partial \rho(\mathbf{x})}{\partial X^i_n}\frac{\partial \rho(\mathbf{y})}{\partial X^j_m} \langle dU_{\varepsilon}(\mathbf{x},t)dU_{\varepsilon}(\mathbf{y},t')\rangle \\
    &= \frac{2k_B T\sigma^{ij}}{\hbar(2\pi\rho_0)^2}\delta(t-t')dt \int d^2\mathbf{x} \frac{\partial \rho(\mathbf{x})}{\partial X^i_n}\left(\int d^2\mathbf{y} \, \frac{\partial \rho(\mathbf{y})}{\partial X^j_m} \varepsilon\left(\mathbf{x}-\mathbf{y}\right)\right) \,. \end{align}\end{subequations}
The integrand of the bracketed integral in Fourier space is a local product of the Fourier transform of $\partial\rho/\partial X^j_m$ and the kernel $\tilde{\varepsilon}(\mathbf{k})$. Following the same argument made in the main text and Section (2) of this document for the energy-damping potential term, we may approximate this integral by noting $\partial\rho/\partial X^j_m$ will be peaked in Fourier space at $k=\xi^{-1}$, allowing us to treat the kernel as constant at this scale: $\tilde{\varepsilon}(\mathbf{k})\approx \tilde{\varepsilon}(\xi^{-1})$. This allows us to make the substitution Eq.~\eqref{eq:approximate_edkernel} in the above correlation,
\begin{subequations}\begin{align}
 \langle dU^i_n(t) dU^j_m(t') \rangle&\approx \sigma^{ij}\frac{4 k_B T \sigma_{\rm ED}N_{\rm cut}}{\hbar (2\pi\rho_0)^2} \delta(t-t')dt\int d^2\mathbf{x}\frac{\partial \rho(\mathbf{x})}{\partial X^i_n} \int d^2\mathbf{y} \, \frac{\partial \rho(\mathbf{y})}{\partial X^j_m}\delta(\mathbf{x-y}) \\
 &= \sigma^{ij}\frac{k_B T \sigma_{\rm ED}N_{\rm cut}}{\pi^2\hbar\rho_0^2} \delta(t-t')dt \int d^2\mathbf{x}  \, \frac{\partial \rho(\mathbf{x})}{\partial X^i_n}\frac{\partial \rho(\mathbf{x})}{\partial X^j_m} \,.
 \end{align}\end{subequations}
The integral can be solved analytically for our choice of density ansatz
\begin{subequations}
	\begin{align}
		\int d^2\mathbf{x}  \, \frac{\partial \rho(\mathbf{x})}{\partial X^i_n}\frac{\partial \rho(\mathbf{x})}{\partial X^j_m} &= \frac{\pi \rho_0^2}{4\xi^2}\exp\left(-\frac{r_{mn}^2}{4\xi^2}\right)\left(2\xi^2\delta_{ij} - \delta X^i_{mn}\delta X^j_{mn}\right) \\
		&\approx \frac{\pi\rho_0^2}{2}\delta_{ij}\delta_{nm} \,,
	\end{align}
\end{subequations}
where in the second line we have again made the approximation $\exp[-r_{mn}^2 / (4 \xi^2)] \approx \delta_{mn}$ valid in the point-vortex regime $r_{mn}^2 \gg \xi^2$. Therefore, off-diagonal correlations vanish, leading to the simple expression:
\begin{subequations}\begin{align}
    \langle dU^i_n(t) dU^j_m(t') \rangle  &= 2\frac{\alpha_\varepsilon k_B T}{2\pi\hbar \rho_0} \delta_{ij}\delta_{nm}\delta(t-t')dt,
 \end{align}\end{subequations}
noting $\sigma^{ij}\delta_{ij}=\delta_{ij}$ and $\sigma_{\rm ED}N_{\rm cut} = 2\alpha_\varepsilon/\rho_0$. This correlation allows us to express the noise vector in terms of white noise processes: 
 \begin{align}
d\mathbf{U}_n(t)\equiv    \sqrt{2\eta}d\mathbf{w}_n(t) = \sqrt{2\eta} \begin{pmatrix}
     dW^x_n(t) \\ dW^y_n(t)
    \end{pmatrix} \,,
\end{align}
where $dW_n^i$ are real Gaussian noises with zero mean and correlation $\langle dW_n^i(t) dW_m^j(t') \rangle = \delta_{ij} \delta_{nm} dt$ (i.e. Weiner increments) and we have defined the diffusion coefficient (units $\text{length}^2/\text{time}$):
\begin{align}
     \eta \equiv\alpha_\varepsilon  \frac{k_B T}{2\pi\hbar\rho_0} \,.
\end{align}
This leads to the stochastic point-vortex equation
\begin{align}
\label{eq:stochasticdPV_withEDSupp}
	d\mathbf{r}_i &= \left(\mathbf{v}^0_i - \alpha_\varepsilon q_i \mathbf{\hat{z}}{\times} \mathbf{v}^0_i\right) dt +\sqrt{2\eta} d\mathbf{w}_n\,,
\end{align}
which is presented as Eq.~(5) of the main text.

\subsection{Numeric Validation: Dipole decay}
\begin{figure}
	\centering
   \includegraphics[width=0.8\columnwidth]{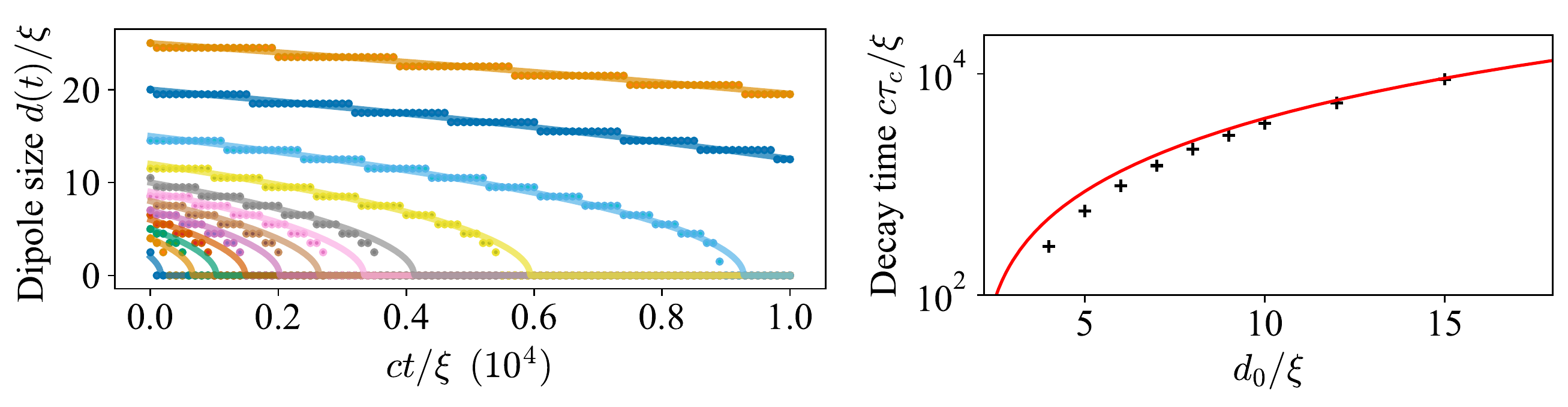}
	\caption{Numerical validation of the damped point-vortex model for dipole decay. (a) Dipole size as a function of time for various initial sizes $d_0\in [2,25]\xi$ as given by the analytic expression Eq.~(\ref{eq:dipole_analytic}) (lines) and numerical integration of the noiseless SPGPE (pluses). (b) Decay time $\tau_c$, defined as $d(\tau_c)=d_c$ for $d_c=2\xi$, given in terms of the speed of sound $c=\sqrt{\mu/m}$, as a function of initial dipole size $d_0\in [4,15]\xi$. Analytic expression Eq.~(\ref{eq:dipole_analytic}) compares well to numeric values, with increasing agreement for $d\gg d_c$. }
   \label{fig:Dipole_ED_AnalyticvsNumeric}
\end{figure}
Here we numerically validate our \zm{choice of density ansatz and} approximate treatment of the energy-damping kernel (i.e. Eq.~\eqref{eq:approximate_edkernel}) by comparing the predictions of our derived point-vortex equation against direct integration of the quasi-2D SPGPE (with the exact expression for the scattering kernel $\varepsilon(\textbf{x})$) for a vertical thickness of $l_z=\xi$. For simplicity, we neglect the noise in both equations, effectively comparing the predictions of each equation for the \emph{mean} vortex dynamics. This allows us to separate deviations due to the approximate form of the kernel \zm{and density} (used in both the damping and noise terms) from sampling errors in averaging over a finite number of stochastic trajectories. 

Specifically, we consider the decay of a vortex-antivortex dipole due to damping, for which the analytic solution to the damped point-vortex equation is well-known (see, for example, [27]).
For our model, this solution can be written as:
\begin{align}
	d(t) = \sqrt{d(0)^2 -4\frac{\hbar}{m}\alpha_\varepsilon t}, \label{eq:dipole_analytic}
\end{align}
where $d(t)$ is the separation between the two vortices and we have neglected the contribution of number damping. By defining a critical scale $d_c$ at which the vortex-antivortex pair are expected to annihilate, we can estimate a timescale for decay $\tau_c=m(d(0)^2-d_c^2)/(4\hbar\alpha_\varepsilon)$. An estimate of this critical scale is $d_c=2\xi$, where $\xi$ is the healing length of the fluid~[27].
Our simulations are performed in dimensionless healing length units of $\xi$ (space) and $c/\xi=\hbar/\mu$ (time), with $l_z=\xi$ and $\mathcal{M}=2\sigma_s N_{\rm cut} \rho_0 = 0.1$ together giving a mutual friction coefficient $\alpha_\varepsilon\approx 0.006$. Here $c=\sqrt{\mu/m}$ is the speed of sound in the superfluid.  \par

Figure \ref{fig:Dipole_ED_AnalyticvsNumeric} compares the analytic expression Eq.~(\ref{eq:dipole_analytic}) to direct numerical integration of the noiseless quasi-2D SPGPE (with $\gamma=0$) for a range of initial dipole sizes. We see strong quantitative agreement between the analytic and numeric results, particularly for large inter-vortex distances $d(t)\gg \xi$. \zm{This clearly demonstrates the validity of the assumptions made in the derivation of our model in the point-vortex limit (e.g. the density ansatz and the approximate form of the kernel), and the stability of our model over long timescales.} We observe growing discrepancy between the analytics for dipole sizes below $d(t)\lesssim5\xi$, indicating the expected breakdown of a point-vortex description of the vortex dynamics. When the approximate form of the energy-damping kernel is also used in the numeric simulations, we observe a slight improvement in the agreement for dipoles with initial separation $d(0)\lesssim10\xi$. This demonstrates that there is a quantitative deviation due to the treatment of the kernel, but one that is small and only noticeable close to the breakdown of the point-vortex regime.

\section{(4) Brownian motion in the mean-field approximation}
Here we show that vortex evolution under the stochastic damped PVM corresponds to Brownian motion in the mean-field limit. Under the mean-field approximation, each vortex interacts with a mean-field flow induced by all other vortices, allowing us to approximate Eq.~\eqref{eq:stochasticdPV_withED} by:
\begin{align}
\begin{pmatrix}
 dX_i\\dY_i
\end{pmatrix} &= \langle \mathbf{v}_i^0 - \alpha_\varepsilon q_i \mathbf{\hat{z}}{\times} \mathbf{v}^0_i\rangle dt - \sqrt{2\eta}\begin{pmatrix}
 dW_i^x\\dW_i^y
\end{pmatrix} \,,
\end{align}
where we have replaced the background superfluid velocity at the $i$-th vortex, $\mathbf{v}_i^0$, with its stochastic average $\langle \mathbf{v}_i^0 \rangle$. In other words, under this assumption the $i$-th vortex interacts with the \emph{mean} velocity field produced by the dynamics of all other vortices.

Using the shorthand $\langle \mathbf{v}_i^0\rangle  - \alpha_\varepsilon q_i \mathbf{\hat{z}}{\times} \langle \mathbf{v}^0_i\rangle \equiv (a_i(t),b_i(t),0)^T$, we can then write the solution of the above equation as a vector Ornstein-Uhlenbeck process:
\begin{subequations}
\begin{align}
    X_i(t) &= x_0 + \int_0^t a_i(t') dt' - \sqrt{2\eta}\int_0^t dW_i^x(t') \,,\\
    Y_i(t) &= y_0 + \int_0^t b_i(t')dt' + \sqrt{2\eta}\int_0^t dW_i^y(t') \,.
\end{align}
\end{subequations}
From here we may then compute the variance of the positions by noting that the noise vanishes in the means ($\langle dW_i^\alpha \rangle = 0$):
\begin{align}
    \langle \Delta X_i^2 \rangle \equiv \left\langle (X_i(t)- \langle X_i(t) \rangle  )^2 \right\rangle &= 2\eta \int_0^t dt' \int_0^t dt'' \langle dW_i^x(t')dW_i^x(t'') \rangle \\
    &= 2\eta\int_0^t dt' = 2\eta t \,.
\end{align}
An identical calculation gives $\langle Y_i^2 \rangle=2\eta t$. Finally, this allows us to compute the growth of the variance induced by thermal fluctuations, in the mean-field approximation:
\begin{align}
    \langle \Delta r_i^2 \rangle \equiv \langle \Delta X_i^2 +\Delta Y_i^2 \rangle = 4\eta t \,.
\end{align}
This can be interpreted as Brownian motion of vortices around the background flow, with diffusive growth of the position variance of each vortex. \par

\section{(5) Estimation of atomic cloud parameters for calculations of the mutual friction coefficient}
 \label{sec:red_2D}
\subsection{Reduction to 2D theory: Calculation of $l_z,\mu_{\rm 2D},\rho_0$ from 3D cloud parameters}
A key parameter in our stochastic point vortex theory is the vertical thickness of the atomic cloud $l_z$. In terms of the quasi-2D SPGPE, this is defined as the $1\sigma$ radius of the transverse wavefunction, which is treated as a Gaussian~[41]. 
In this work we compute $l_z$ for a given harmonically trapped system based on the analytical variational ground state for a Gaussian ansatz, as given in Ref.~[46]. 
Specifically, we find $l_z$ as the solution to the following algebraic equation ($b_i = l_i/\sqrt{\hbar/(m\omega_i)} $):
\begin{align}
	  \frac{1}{2}\hbar\omega_i\left(b_i^2 - \frac{1}{b_i^2}\right) - \frac{1}{2(2\pi)^{3/2}}\frac{gN_0}{l_{\rm geo}^3}\frac{1}{b_1b_2b_3}=0
\end{align}
where $\omega_{\rm geo}=(\omega_x\omega_y\omega_z)^{1/3}$ is the geometric mean of the trapping frequencies, $l_{\rm geo}=\sqrt{\hbar/(m\omega_{\rm geo})}$, and $N_0$ is the number of condensate atoms.

Integrating out the $z$ dimension results in an effective 2D chemical potential and interaction strength:
\begin{align}
	\mu_{\rm 2D} &= \mu - \frac{m\omega_z^2l_z^2 }{4}- \frac{\hbar^2}{4ml_z^2} \,, \\
	g_{\rm 2D} &= \frac{g}{\sqrt{2\pi}l_z} \,,
\end{align}
which we use to estimate the healing length $\xi=\hbar/\sqrt{m\mu_{\rm 2D}}$ and the 2D background density $\rho_0=\mu_{\rm 2D}/g_{\rm 2D}$.

\subsection{Estimation of chemical potential for comparison to ZNG simulations of Ref.~[47]}

In the main text we compare our microscopically-derived expression for the mutual friction coefficient to the numerical calculations of Ref.~[47]. 
In their calculations, the total atom number $N_T$ of the gas was fixed for all temperatures studied, resulting in a different chemical potential for each temperature considered -- therefore changing the effective energy cutoff for each temperature. For each temperature $T$, we compute the chemical potential by first estimating the number of condensate atoms $N_0$, using the thermodynamic expression~[45]:
\begin{align}
    \frac{N_0}{N_T} = \left[1-\left(\frac{T}{T_c^0}\right)^3\right] -\frac{3\omega_{\rm arith} \zeta(2) }{2\omega_{\rm geo} [\zeta(3)]^{2/3}}\left(\frac{T}{T_c^0}\right)^2 N_T^{-1/3}  \,,
\end{align}
where $\omega_{\rm arith}=(\omega_x+\omega_y+\omega_z)/3$ is the arithmetic mean of the trapping frequencies, and $T_c^0=177$~nK is the ideal gas critical temperature for the parameters in Ref.~[47]. 
The first term in this equation is simply the ideal-gas relation, and the second accounts for the first-order shift in the critical temperature due to the finite-size of the trapped gas. From $N_0$, the chemical potential can then be estimated in the Thomas-Fermi approximation~[45]:
\begin{align}
    \mu = \frac{\hbar \omega_{\rm geo}}{2}\left(\frac{15 N_0 a_s}{l_{\rm geo}}\right)^{2/5} \,.
\end{align}
This value of $\mu$ is then used to set the energy cutoff $\epsilon_{\rm cut}=2\mu$ and compute the 2D background density as described above.

{\small
$^*$~zain.mehdi@anu.edu.au
\begin{enumerate}
	\setlength\itemsep{0.05em}
	\setcounter{enumi}{4}
	\item  S.~J.~Rooney, T.~W.~Neely, B.~P.~Anderson, and A.~S.~Bradley, Phys. Rev. A \textbf{88}, 063620 (2013) 

	\setcounter{enumi}{23}
	\item  P.~Blakie, A.~Bradley, M. Davis, R. Ballagh, and C.~Gardiner, Advances in Physics \textbf{57}, 363 (2008).
	\setcounter{enumi}{25}
	\item G.~Moon, W.~J.~Kwon, H.~Lee, and Y.-i.~Shin, Physical Review A \textbf{92}, 051601 (2015).	\setcounter{enumi}{26}
	\item W.~J.~Kwon, G.~Del~Pace, K.~Xhani, L.~Galantucci, A.~Muzi Falconi, M.~Inguscio, F.~Scazza, and G.~Roati, Nature 600, \textbf{64} (2021).
	\setcounter{enumi}{28}
	\item S.~J.~Rooney, A.~S.~Bradley, and P.~B.~Blakie, Physical Review A \textbf{81}, 023630 (2010).
	\setcounter{enumi}{30}
	\item A.~S.~Bradley, C.~W.~Gardiner, and M.~J.~Davis,
Physical Review A \textbf{77}, 033616 (2008)
	\setcounter{enumi}{31}
	\item A.~S.~Bradley and B.~P.~Anderson, Physical Review X \textbf{2}, 041001 (2012).
	\setcounter{enumi}{40} 
	\item A.~S.~Bradley, S.~J.~Rooney, and R.~G.~McDonald, Physical Review A \textbf{92}, 033631 (2015)
	\setcounter{enumi}{41} 
	\item Z.~Mehdi, A.~S.~Bradley, J.~J.~Hope, and S.~S.~Szigeti, SciPost Phys. \textbf{11}, 80 (2021).	\setcounter{enumi}{43}
	\item A.~Lucas and P.~Surowka, Physical Review A \textbf{90}, 053617 (2014).
	\setcounter{enumi}{44}
	\item  F.~Dalfovo, S.~Giorgini, L.~P.~Pitaevskii, and
S.~Stringari, Rev. Mod. Phys. \textbf{71}, 463 (1999).
	\setcounter{enumi}{45}
	\item C.~J.~Pethick and H.~Smith, ``Bose–Einstein Condensation in Dilute Gases'' (Cambridge University Press, 2008).
	\setcounter{enumi}{46}
	\item B.~Jackson, N.~P.~Proukakis, C.~F.~Barenghi, and E.~Zaremba, Physical Review A \textbf{79}, 053615 (2009)
	\setcounter{enumi}{47}
	\item  S.~J.~Rooney, P.~B.~Blakie, B.~P.~Anderson and A.~S.~Bradley, Physical Review A \textbf{84}, 023637 (2011).
\end{enumerate}
}

\end{document}